\documentstyle[twoside,fleqn,espcrc2]{article}
 
\newcommand{\Tr}{{\rm Tr}}
\input epsf
\def\m@th{\mathsurround=0pt}
\def\eqalign#1{\null\,\vcenter{\openup\jot \m@th
  \ialign{\strut\hfil$\displaystyle{##}$&$\displaystyle{{}##}$\hfil
      \crcr#1\crcr}}\,}

\newcommand{\AmS}{{\protect\the\textfont2
  A\kern-.1667em\lower.5ex\hbox{M}\kern-.125emS}}
 
\title{ Species Doubling and Effective Lagrangians}

\author{Michael Creutz and Michel Tytgat
\address{Physics Department, Brookhaven
 National Laboratory, PO Box 5000, Upton, NY 11973-5000, USA\\
 creutz@wind.phy.bnl.gov,mtytgat@wind.phy.bnl.gov
}
\thanks{Poster presented by M. Creutz.
This manuscript has been authored under contract number
DE-AC02-76CH00016 with the U.S.~Department of Energy.  Accordingly,
the U.S.~Government retains a non-exclusive, royalty-free license to
publish or reproduce the published form of this contribution, or allow
others to do so, for U.S.~Government purposes.}
}

\begin{document}
 
\begin{abstract}
Coupling gauge fields to the chiral currents from an effective
Lagrangian for pseudoscalar mesons naturally gives rise to a species
doubling phenomenon similar to that seen with fermionic fields in
lattice gauge theory.
\end{abstract}
 
\maketitle

Species doubling is deeply entwined with the famous axial anomalies.
A lattice regulator removes all infinities; so, anomalies cannot
appear without explicit symmetry breaking.  As the regulator is
removed, a remnant remains determining an overall chiral phase.  This
phase is the well known strong CP parameter $\theta$ \cite{theta}.
Most lattice schemes also mutilate non-singlet chiral symmetries,
although ways around this exist \cite{shamir}.

For the gauge interactions of the electroweak theory, non-perturbative
chiral issues remain unresolved.  The $W$ bosons couple in a parity
violating manner, and rely for consistency on a subtle anomaly
cancellation between quarks and leptons.  While a flurry of recent
work treats fermions with a separate limit
\cite{twocutoff,overlap2,lee}, it remains unclear how to implement
this cancellation in a fully finite and gauge-invariant manner.

We argue that the doubling problem is not unique to the lattice
approach, but is a general consequence of chiral anomalies.  This
presentation is based on our recent letter \cite{mcmt}.  Starting with
an effective Lagrangian for the pseudoscalar mesons, we couple gauge
fields to the chiral currents.  When these fields are chiral, the
process naturally introduces additional particles mirroring the
original theory.  As with lattice doublers, the mirror fields cancel
anomalies.

In a chiral Lagrangian approach, the effects of anomalies appear in a
term introduced by Wess and Zumino \cite{wz}, and later elucidated by
Witten \cite{witten}.  This involves extending the fields into an
internal space, only the boundary of which is relevant to the
equations of motion.  On coupling to a local gauge field, the boundary
can acquire additional contributions.  We argue that these are most
naturally written in terms of doubler fields.

We start with quark fields $\psi^a(x)$ interacting with
non-Abelian gluons.  We suppress all indices except flavor,
represented by the index $a$, and space-time, represented by $x$.
From $\psi$ we project right and left handed parts, $\psi^a_R={1\over
2} (1+\gamma_5)\psi^a$ and $\psi^a_L={1\over 2} (1-\gamma_5)\psi^a$.
We ignore fermion masses.

This theory with massless quarks is invariant under a global
$SU(n_f)\times SU(n_f)$ symmetry, where $n_f$ represents the number of
flavors.  The quark fields transform as $\psi_L^a\rightarrow \psi_L^b
g_L^{ba}$ and $\psi_R^a\rightarrow \psi_R^b g_R^{ba}$.  Here $g_L$ and
$g_R$ are elements of $SU(n_f)$.

The chiral symmetry is spontaneously broken by the {\ae}ther, resulting
in $n_f^2-1$ Goldstone bosons and a remaining explicit $SU(n_f)$
flavor symmetry.  The composite field $\overline \psi^a_R \psi^b_L$
acquires an expectation value.  Chiral symmetry allows chosing a
standard {\ae}ther with, say, $\langle\overline \psi^a_R
\psi^b_L\rangle=v\delta^{ab}$.  Here the parameter $v$ determining the
magnitude of the expectation value requires a renormalization scheme
for precise definition.  The {\ae}ther is degenerate, and one could
choose to replace $\delta^{ab}$ by an arbitrary element $g^{ab}$ of
$SU(n_f)$.  The basic idea of the effective Lagrangian is to promote
this element into a local field $g(x)$.  Slow variations of this field
represent the Goldstone bosons.  The chiral symmetry is an invariance
under $g(x)\rightarrow g_L^\dagger g(x) g_R$.

The effective Lagrangian approach represents an expansion in light
particle momenta \cite{clreview}.  The lowest order action is
\begin{equation}
S_0={F_\pi^2\over 4}\int d^4x\ {\rm Tr}(\partial_\mu g\partial_\mu g^\dagger). 
\end{equation}
The constant $F_\pi$ has an experimental
value 93 MeV.  In terms of conventional
fields, $g=\exp(i\pi\cdot\lambda/F_\pi)$, where the
$n_f^2-1$ matrices $\lambda$ generate $SU(n_f)$ and are normalized ${\rm Tr}
\lambda^\alpha \lambda^\beta=2\delta^{\alpha\beta}$.

From this action the equations of motion are $\partial_\mu
J_{L,\mu}^\alpha=0$, where the ``left'' current is 
\begin{equation}
J_{L,\mu}^\alpha= {iF_\pi^2\over 4}{\rm Tr}\lambda^\alpha(\partial_\mu g)g^\dagger
\end{equation}
Equivalently, one can work with ``right'' currents.  There is a vast
literature on adding higher derivative terms \cite{clreview}.

We are interested in a special higher derivative coupling describing
the effects of anomalies.  This term cannot be written as an integral
of a local expression in $g(x)$, even though the resulting
contribution to the equations of motion is fully local
\cite{wz,witten,zumino}.  Continuing to write the equations of motion
in terms of a divergence free current, a possible addition which
satisfies the required symmetries is
\begin{equation}
\eqalign{
J_{L,\mu}^\alpha&=
{iF_\pi^2\over 4}{\rm Tr}\lambda^\alpha(\partial_\mu g)g^\dagger\cr
&+{in_c\over 48\pi^2} \epsilon_{\mu\nu\rho\sigma}{\rm Tr}\lambda^\alpha
(\partial_\nu g)g^\dagger
(\partial_\rho g)g^\dagger
(\partial_\sigma g)g^\dagger\cr
}
\end{equation}

To obtain an action generating the above requires extending $g(x)$
beyond a simple mapping of space-time into the group.  We introduce an
auxiliary variable $s$ to interpolate between the field $g(x)$ and
some fixed group element $g_0$.  Thus consider $h(x,s)$ satisfying
$h(x,1)=g(x)$ and $h(x,0)=g_0$.  This extension is not unique, but the
equations of motion are independent of the chosen path.  We now write
\begin{equation}
\eqalign{
S&={F_\pi^2\over 4}\int d^4x\ 
{\rm Tr}(\partial_\mu g\partial_\mu g^\dagger)\cr
&+{n_c\over 240 \pi^2}\int d^4x\int_0^1 
ds\ \epsilon_{\alpha\beta\gamma\delta\rho}
{\rm Tr} h_\alpha h_\beta h_\gamma h_\delta h_\rho.\cr
}
\end{equation}
Here we define $h_\alpha=i(\partial_\alpha h)
h^\dagger$ and regard $s$ as a fifth coordinate.  

For equations of motion, consider a small variation of $h(x,s)$.  This
changes the final integrand by a total divergence, which then
integrates to a surface term.  Working with either spherical or
toroidal boundary conditions in the space-time directions, this
surface only involves the boundaries of the $s$ integration.  When
$s=0$, space-time derivatives acting on the constant matrix $g_0$
vanish.  The surface at $s=1$ generates precisely the desired
additional term in Eq.~(3).

The last term in Eq.~(4) represents a piece cut from the $S_5$ sphere
appearing in the structure of $SU(n_f)$ for $n_f\ge 3$.  The mapping
of four dimensional space-time into the group surrounds this volume.
Chiral rotations shift this region around, leaving its measure
invariant.  As emphasized by Witten \cite{witten}, this term is
ambiguous.  Different extensions into the $s$ coordinate can modify
the above five dimensional integral by an integer multiple of
$480\pi^3$.  To have a well defined quantum theory, the action must be
determined up to a multiple of $2\pi$.  Thus the quantization of $n_c$
to an integer, the number of ``colors.''

Crucial here is the irrelevance of the starting group element $g_0$
at the lower end of the $s$ integration.  Our main point is the
difficulty of maintaining this condition when we make the chiral
symmetry local.  As usual, this requires the introduction of gauge
fields.  Under the transformation $g(x)\rightarrow g_L^\dagger(x) g(x)
g_R(x)$, derivatives of $g$ transform as
\begin{equation}
\partial_\mu g\longrightarrow
g_L^\dagger\left (
\partial_\mu g-\partial_\mu g_L g_L^\dagger g+ g\partial_\mu g_R g_R^\dagger 
\right) g_R
\end{equation}
To compensate, we introduce left and right gauge fields transforming as
\begin{equation}\matrix{
A_{L,\mu} & \longrightarrow
& g_L^\dagger A_{L,\mu} g_L + i g_L^\dagger\partial_\mu g_L \cr
A_{R,\mu} & \longrightarrow 
& g_R^\dagger A_{R,\mu} g_R + i g_R^\dagger\partial_\mu g_R \cr
}
\end{equation}
Then the combination
\begin{equation}
D_\mu g = \partial_\mu g-iA_{L,\mu} g + ig A_{R,\mu}
\end{equation} 
transforms nicely: $D_\mu g\rightarrow g_L^\dagger D_\mu g g_R$.  Making
the generalized minimal replacement $\partial_\mu g\rightarrow
D_\mu g$ in $S_0$, we find a gauge invariant action.

A problem arises when we go on to the Wess-Zumino term.  We require a
prescription for the gauge transformation on the interpolated group
element $h(x,s)$.  Here we note a striking analogy with the domain
wall approach to chiral fermions first promoted by Kaplan
\cite{kaplan}.  There an extra dimension was also introduced, with the
fermions being surface modes bound to a four dimensional interface.
The usual approach to adding gauge fields involves, first, not giving
the gauge fields a dependence on the extra coordinate, and, second,
forcing the component of the gauge field pointing in the extra
dimension to vanish \cite{mcih,gjk}.  In terms of a five dimensional
gauge field, we take $A_\mu(x,s)=A_\mu(x)$ and $A_s=0$ for both the
left and right handed parts.  Relaxing either of these would introduce
unwanted degrees of freedom.  The natural extension of the gauge
transformation is to take $h(x,s)\rightarrow g_L^\dagger(x) h(x,s)
g_R(x)$ with $g_{L,R}$ independent of $s$.

We now replace the derivatives in the Wess-Zumino term with covariant
derivatives.  This alone does not give equations of motion independent
of the interpolation into the extra dimension.  However, adding terms
linear and quadratic in the gauge field strengths allows construction
of a five dimensional Wess-Zumino term for which variations are again
a total derivative.  This gives
\begin{equation}
S_{WZ}=
{n_c\over 240 \pi^2 }\int d^4x \int_0^1 ds\  \Gamma
\end{equation}
where
\begin{equation}\eqalign{
\Gamma&=\Gamma_0+{5i\over2}(i\Gamma_L+i\Gamma_R\cr
&-\Gamma_{LL}-\Gamma_{RR}-\alpha\Gamma_{LR}-(1-\alpha)\Gamma_{RL}),\cr
}
\end{equation}
$\alpha$ is a free parameter, and 
\begin{equation}\eqalign{
&\Gamma_0=
\epsilon_{\mu\nu\rho\lambda\sigma} 
\Tr D_\mu h h^\dagger D_\nu h h^\dagger D_\rho h h^\dagger 
D_\lambda h h^\dagger D_\sigma h h^\dagger\cr
&\Gamma_L=\epsilon_{\mu\nu\rho\lambda\sigma} 
\Tr D_\mu h h^\dagger D_\nu h h^\dagger D_\rho h h^\dagger
F_{L,\lambda\sigma} \cr
&\Gamma_R=\epsilon_{\mu\nu\rho\lambda\sigma} 
\Tr D_\mu h h^\dagger D_\nu h h^\dagger D_\rho h
F_{R,\lambda\sigma} h^\dagger \cr
&\Gamma_{LL}=
\epsilon_{\mu\nu\rho\lambda\sigma} 
\Tr D_\mu h h^\dagger F_{L,\nu\rho} F_{L,\lambda\sigma} \cr
&\Gamma_{RR}=
\epsilon_{\mu\nu\rho\lambda\sigma} 
\Tr D_\mu h  F_{R,\nu\rho} F_{R,\lambda\sigma} h^\dagger \cr
&\Gamma_{RL}=
\epsilon_{\mu\nu\rho\lambda\sigma} 
\Tr D_\mu h  F_{R,\nu\rho} h^\dagger F_{L,\lambda\sigma} \cr
&\Gamma_{LR}=
\epsilon_{\mu\nu\rho\lambda\sigma} 
\Tr D_\mu h h^\dagger F_{L,\nu\rho} h F_{R,\lambda\sigma} h^\dagger.\cr
}
\end{equation}
The covariantly transforming field strengths are
\begin{equation}
\eqalign
{
&F_{L,\mu\nu}=\partial_\mu A_{L,\nu}-\partial_\nu A_{L,\mu}
-i[A_{L,\mu},A_{L,\nu}]\cr
&F_{R,\mu\nu}=\partial_\mu A_{R,\nu}-\partial_\nu A_{R,\mu}
-i[A_{R,\mu},A_{R,\nu}].\cr
}
\end{equation}
For the photon, parity invariance fixes $\alpha=1/2$.  The
last four terms contain the process $\pi\rightarrow 2\gamma$.

This procedure works well for a vector-like gauge field, where we take
$g_L(x)=g_R(x)$ and $A_L=A_R$.  We could, for example, take $g_0$ to be
the identity, and then the gauge transformation cancels out at $s=0$.
However, difficulties arise on
coupling a gauge field to an axial current.
Then $g_0\rightarrow g_L^\dagger(x) g_0 g_R(x)$ in general
will no longer be a constant group element.  After a gauge
transformation, variations of the action give new non-vanishing
contributions to the equations of motion from the lower end of the $s$
integration.

The simplest solution makes the $s=0$ fields dynamical.  Thus
we replace the field $g(x)$ with two fields $g_0(x)$ and $g_1(x)$.  The
interpolating field now has the properties $h(x,0)=g_0(x)$ and
$h(x,1)=g_1(x)$.  The action becomes
\begin{equation}
{F_\pi^2\over 4}
\int d^4x\ {\rm Tr}(D_\mu g_0D_\mu g_0^\dagger+D_\mu g_1D_\mu g_1^\dagger)
+ S_{WZ}.
\end{equation}

While now gauge invariant, the theory differs from the
starting model through a doubling of meson species.  The extra
particles are associated with the second set of group valued fields
$g_0(x)$.  The Wess-Zumino term of the new fields has the
opposite sign since it comes from the lower end of the $s$
integration.  Thus, these ``mirror'' particles have reflected chiral
properties and implement a cancellation of all anomalies.  In essence,
we have circumvented the subtleties in gauging the model.
The value of $F_\pi$ need not be the same for $g_0$ and $g_1$; so,
their strong interactions might differ in scale.  Nevertheless,
coupling with equal magnitude to the gauge bosons, the new fields
cannot be ignored.  

For vector currents, we can remove the doublers using a diagonal mass
term at $s=0$.  For example, with a term $M{\rm Tr} g_0(x)$ added to
the Lagrangian density, $M$ could be arbitrarily large, forcing $g_0$
towards the identity.

The doublers arise in analogy to the problems appearing in the surface
mode approach to chiral lattice fermions
\cite{kaplan,mcih,gjk,goltermanshamir}.  In both cases, an extension
to an extra dimension is introduced.  Difficulties arise from the
appearance of an extra interface.  This new surface couples with equal
strength to the gauge fields.

If we let $g_{L,R}$ depend on $s$, we expect problems similar to those
seen with domain wall fermions.  When the gauge fields vary in the
extra dimension, four dimensional gauge invariance is lost.  Symmetry
can be restored via a Higgs field, but this introduces the possibility
of unwanted degrees of freedom in the physical spectrum.
Ref.\cite{gjk} explores the possibility of sharply truncating the
gauge field at an intermediate value of the extra coordinate.  This
gives rise to new low energy bound states acting much like the
undesired doubler states.

A Higgs field does permit different masses for the extra species.  In
particular, the matter couplings to the Higgs field can depend on $s$.
Qualitative arguments suggest that triviality effects on such
couplings limit their strength, precluding masses for the extra
species beyond a typical weak interaction scale.  Presumably such
constraints are strongest when the anomalies in the undoubled sector
are not canceled.  With domain-wall fermions, taking a Higgs-fermion
coupling to infinity on one wall introduces a plethora of new low
energy bound states \cite{goltermanshamir}.

These problems emphasize the subtle way anomalies cancel between
quarks and leptons.  If the contributions of the leptons are ignored,
no non-perturbative approach can be expected to accommodate gauged
weak currents.  When the required cancellations occur between
different fermion representations, perturbation theory appears to be
consistent, while all known non-perturbative approaches remain
awkward.

There are several possible solutions.  Mirror particles might exist,
perhaps with masses at the weak scale \cite{montvay}.  Such might even
be useful in the spontaneous breaking of the electroweak theory
\cite{technicolor}.  A related alternative involves spontaneous
breaking of an underlying vector-like theory containing additional
heavy bosons coupling with opposite parity fermions \cite{patisalam}.
All of these involve a profusion of new particles awaiting discovery.
A speculative solution would twist the extra dimension so that the
doubling particles could be among those already observed.  This
requires the interpolation in the extra dimension to mix the quarks
and the leptons, all of which are involved in the anomaly
cancellations.

\end{document}